\documentclass[12pt]{JHEP3}
\usepackage{amsmath,epsfig}
\def\be{\begin{equation}}
\def\ee{\end{equation}}
\def\ba{\begin{eqnarray}}
\def\ea{\end{eqnarray}}

\def\beq{\begin{equation}}
\def\eeq{\end{equation}}
\def\beqa{\begin{eqnarray}}
\def\eeqa{\end{eqnarray}}

\def\yzero{\smash{\hbox{$y\kern-4pt\raise1pt\hbox{${}^\circ$}$}}}

\def\beq{\begin{equation}}
\def\eeq{\end{equation}}
\def\beqa{\begin{eqnarray}}
\def\eeqa{\end{eqnarray}}

\def\-{\hphantom{-}}

\def\s2{\frac{1}{\sqrt2}}

\def\beq{\begin{equation}}
\def\eeq{\end{equation}}
\def\beqa{\begin{eqnarray}}
\def\eeqa{\end{eqnarray}}

\def\IF{\relax{\rm I\kern-.18em F}}
\def\II{\relax{\rm I\kern-.18em I}}
\def\IP{\relax{\rm I\kern-.18em P}}
\def\IC{\relax\hbox{\kern.25em$\inbar\kern-.3em{\rm C}$}}
\def\IR{\relax{\rm I\kern-.18em R}}

\def\Dsl{\,\raise.15ex\hbox{/}\mkern-13.5mu D} %this one can be subscripted
\def\IZ{Z\kern-.4em  Z}

\title{The Supermembrane with Central Charges on a G2 Manifold}

\author{A. Belhaj$^1$,
  M.P. Garcia del Moral$^2$, A. Restuccia$^{3,4}$, A. Segui$^1$,
  J.P. Veiro$^5$,\footnote{E-mail: \emph{belhaj@unizar.es; garcia@to.infn.it;
restucci@aei.mpg.de, arestu@usb.ve;\newline segui@unizar.es;
pierre@ma.usb.ve}}\\
$^1$ Departamento de F\'\i sica Te\'orica, Universidad de Zaragoza\\
12, Pedro Cerbuna, 50009-Zaragoza, Spain.\\
 $^2$ Dipartimento di Fisica Teorica, Universit\`a di Torino\\
 $\&$ INFN - Sezione di Torino; Via P. Giuria 1; I-10125 Torino, Italy.\\
 $^3$ Max-Planck-Institut f\"ur Gravitationphysik, Albert-Einstein-Institut\\
 M\"ulenberg 1, D-14476 Potsdam, Germany.\\
 $^4$ Departamento de F\'\i sica, Universidad Sim\'on Bol\'\i var\\
Apartado 89000, Caracas 1080-A, Venezuela. \\
$^5$ Departamento de Matem\'aticas Puras y Aplicadas, Universidad Sim\'on Bol\'\i var\\
Apartado 89000, Caracas 1080-A, Venezuela.}

\abstract{We construct the 11D supermembrane with topological
central charges induced through an
 irreducible winding on a G2 manifold rea\-lized from the $T^{7}/Z_{2}^{3}$
 orbifold construction.
 The hamiltonian $H$ of the theory on a $T^{7}$ target has a discrete spectrum.
 Within the discrete symmetries of $H$ associated to large diffeomorphisms,
 the $Z_{2}\times Z_{2}\times Z_{2}$ group of automorphisms of the
 quaternionic subspaces preserving the octonionic
 structure is relevant. By performing the corresponding identification on the
 target space, the supermembrane may be formulated on a G2 manifold,
 preserving the discretness of its supersymmetric spectrum.
The corresponding 4D low energy effective field
 theory has $N=1$ supersymmetry.}

\preprint{DFTT-05/2008\\
AEI-2008-014}
\keywords{Supermembrane, minimal immersions, G2 manifolds,
supersymmetric spectrum}

\begin{document}

\section{Introduction}
Compactifications of the low energy limit of M-theory to four dimensions (4D)
 have received much attention during the past years. Special interest has been
given to the compactification over real manifolds of dimension seven, X7,
with non trivial holonomy. This interest is due to the fact that these manifolds
provide a potential point of contact with low energy semi-realistic physics from M-theory~\cite{TP,Ach}.\\

\paragraph{Our goal} along this paper will be to show the quantization of
the supersymmetric action of the supermembrane, restricted by a
topological condition, on a particular G2 manifold. This is the only
(known) quantum consistent way of doing it starting from the
supermembrane, there is no other way so far. We will show that this
model represents a starting point of a new kind of
\emph{supersymmetric quantum consistent} models with potentially
interesting properties from a phenomenological point of
view.\newline

Let us make a brief review of the main properties of the compactification of M-theory on G2 manifolds.
In particular, one can obtain 4D $N = 1$ supersymmetry by compactifying M-theory on X7 with G2 holonomy group
~\cite{HM,KM,AW}. In this regard, the 4D $N = 1$ resulting models generically depend on geometric properties of X7.
For instance, if X7 is smooth, the low energy theory contains, in addition to $N = 1$ supergravity,
only abelian gauge groups and neutral chiral multiplets. However,
non abelian gauge symmetries with charged chiral fermions
 can be obtained by considering limits where X7 develops singularities~\cite{W,AcW,BDR,ADIL, ABLY}. For a review see for example \cite{AG}.
\\

Besides the ordinary compactification on G2 manifolds, Calabi-Yau flux compactifications
and twisted toroidal compactifications have been also studied intensively,
 see for example,~\cite{CCQ}-~\cite{DF}. Indeed their respectively phenomenological
 predictions with different signatures on the LHC have also been considered, see \cite{ABGKV, ABKKS}
 for G2 compactifications, and \cite{CKSAQ} for large volume approach in Calabi-Yau compactifications.
 They have also been considered as particular cases of
 non-geometric compactifications. Most of these approaches follow a bottom-up
 pattern by studying the $N=1$ gauged supergravity potentials in 4D and trying to
 perform the uplift to M-theory. Other compactifications from 11D supergravity with
 fluxes have also been done in a top-down approach~\cite{DAF,DF}.
\\

Recently new types of compactifications have appeared involving twisted boundary
conditions or non-trivial fiber bundles over some compact manifolds (with or without singularities),
 T- foldings \cite{hull}. In this way, the metric and the gauge field forms get generically entangled.
  This kind of compactifications is called non-geometric
~\cite{DbH, wecht}.
  Some of these non-geometric compactifications are related with the ordinary ones by dualities.
  The nontriviality of the fiber bundle guarantees the existence of a monodromy, but usually due
  to the lack of 1-cycles inside a Calabi-Yau makes it necessary to include singularities.
  A simple example of these T-foldings is the twisted tori. It is a Scherk-Schwarz compactification
  of the 11D supergravity theory  with twisted boundary conditions that allows to have a nontrivial
  monodromy, see for example in connection with G2 compactifications \cite{agatha-prezas}. When the base space is a torus it
is no longer necessary to include singularities
  in order to have a nontrivial monodromy~\cite{DbH,wecht}.
These twisted compactifications can have
  a geometrical dual which corresponds to an orbifold plus a shift,
also known as asymmetric orbifold \cite{blumenhagen}.
\\

The compactification with a duality twist is more general than the
orbifold compactification because it can be carried without restricting the moduli
to special variations. The moduli can have nontrivial variation along the circle in
the spacetime. However, the orbifold is possible for special values of the moduli where
the lattice admits a symmetry and the class of allowed rotations is finite. All of the
lattices admit a $Z_{2}$ symmetry as the discrete subgroup of the $SL(2,Z)$ of the torus,
and for those cases the geometrical dual exists \cite{hull2}.
\\

The 11D supermembrane is one of the basic elements of M-theory
~\cite{bst, dwhn}. Classically, it is unstable due to the existence of
string-like spikes that leave the energy unchanged. At the quantum
level, its supersymmetric spectrum is continuous and the theory was
interpreted as a second quantized theory \cite{dwln, dwmn}. Compactification on
$S^{1}$ has been explored in order to see if the continuity of the
spectrum is broken by the winding. It has been argued not to be the
case~\cite{dwpp} due to the presence of string-like spikes in the
spectrum. In~\cite{torrealba, mor,gmr,bgmmr,bgmr, bellorin} the
minimally immersed supermembrane compactified on a torus associated
to the existence of irreducible winding (MIM2) has been found. It is
associated to nontrivial fiber bundles defined on Riemann surfaces.
This MIM2 is classically stable since there are no singular
configurations with zero energy. The quantum spectrum of the theory
is purely discrete with finite multiplicity~\cite{gmr,bgmmr,bgmr,br,bgmr2}.
 The theory of the supermembrane minimally immersed in a 7 torus
has recently been found in~\cite{joselen}. It has a $N=1$
supersymmetry in 4D. A natural question is to look for a connection
with a compactification of the supermembrane in a nontrivial
background of type of G2 manifold. In this paper we will be
concerned with a full-flegged sector of M-theory which is the
quantum supermembrane theory minimally immersed MIM2 on a $T^{6}\times S^{1}$.
This type of compactifications contains nontrivial discrete twists on the fibers as
remanent discrete symmetries of the hamiltonian. We will show by identifying those symmetries on the target
that the MIM2 can admit a compactification on a G2 manifold.
\\

The paper is structured as follows: In Section~2 we introduce the
supermembrane with central charges minimally immersed (M2MI) on a
$T^{6}\times S^{1}$. We summarize its main spectral properties and
symmetries. In Section~3 we recall the main properties of the
compactification on G2 manifolds. In Section~4 we construct the MIM2
on a $\frac{T^{7}}{Z_{2}^{3}}$, by studying the minimal immersions
of the MIM2 on that target, finding the configuration space of
states: the untwisted and twisted sectors of the theory. We also
study its connections with Calabi-Yau compactifications. In
Section~5 we present our discussions and final conclusions.
\section{M2 with central charges associated to an irreducible winding}
We start this section by recalling that the hamiltonian of the $D=11$ supermembrane~\cite{bst} in the light cone gauge (LCG) reads as
\begin{equation}\label{ha}
\int_\Sigma \sqrt{W} \left(\frac{1}{2}\left(\frac{P_M}{\sqrt{W}}\right)^2+\frac{1}{4} \{X^M,X^N\}^2+{\small\mathrm{\ Fermionic\ terms\ }}\right).
\end{equation}
$M$ runs for $M=1,\dots,9$ corresponding to the transverse coordinates of the base manifold $R\times \Sigma$. $\Sigma$ is a Riemann surface of genus $g$. The term $\frac{P_{M}}{\sqrt{W}}$ is the canonical momentum density and $ \{X^{M}, X^{N}\}$ is given by
\be \label{e4}
\{X^{M}, X^{N}\}= \frac{\epsilon^{ab}}{\sqrt{W(\sigma)}}\partial_{a}X^{M}\partial_{b}X^{N},
\ee
where $a,b=1,2$ and $\sigma^{a}$ are local coordinates over $\Sigma$. $W(\sigma)$ is a scalar density introduced in the LCG fixing procedure. The former hamiltonian is subject to the two following constraints
\begin{equation}
\label{e2} \phi_{1}:=d(\frac{P_{M}}{\sqrt{W}}dX^{M})=0
\end{equation}
\begin{equation} \label{e3}
\phi_{2}:= \oint_{C_{s}}\frac{P_M}{\sqrt{W}} \quad dX^M = 0,
\end{equation}
where $C_{s}, s=1,\dots,2g$ is a basis of 1-dimensional cycles on $\Sigma$. $\phi_{1}$ and $\phi_{2}$ are generators of area preserving diffeomorphisms. When the target manifold is simply connected, the one-forms $dX^{M}$ are exact.
\\

The $SU(N)$ regularized model obtained from (\ref{ha}) was shown to have continuous
spectrum from $[0,\infty)$,~\cite{dwln, dwmn, dwhn}. This property of the theory relies
on two basic facts: supersymmetry and the presence of classical singular configurations.
The latter is related to string-like spikes which appear with zero cost energy.
These spikes do not preserve neither the topology of the world-volume nor the number of particles.
These properties do not disappear when the theory is compactified and the spectrum remains continuous~\cite{dwpp}.
\\

To get a 4 dimensional model, we need a target space as $M_{4}\times T^{6}\times S^{1}$. In this way,
the configuration maps satisfy the following condition on $T^{6}$
\begin{equation}\label{e5}
\oint_{c_{s}}dX^{r}=2\pi S_{s}^{r}R^{r}\quad r,s=1,\dots,6.
\end{equation}
On the circle, we have the constraint
\begin{equation}\label{e6}
\oint_{c_{s}}dX^{7}=2\pi L_{s}R_7
\end{equation}
while for non compact directions, we have
\begin{equation}\label{e7}
\oint_{c_{s}}dX^{m}=0 \quad m=8,9.
\end{equation}
$S_{s}^{r},L_{s}\in Z $ and $R_{r},R_{7}$ represent respectively the radii of the 6-torus
and the radius of the circle. We shall now impose a topological irreducible wrapping condition
to be satisfied by all configurations in the above model. This generates a non-trivial central charge
in the 11D supersymmetric algebra. The topological condition is
\be\label{e8}
I^{rs}\equiv\int_{\Sigma}dX^{r}\wedge dX^{s}=n(2\pi R^{r}R^{s})\omega^{rs}
\ee
where $\omega^{rs}$ is a symplectic matrix on the $T^{6}$ which can be taken as
\begin{equation}\label{matrix}
\omega^{rs} =\begin{pmatrix}
0 & 1 & & & &\\
-1 & 0 & & & & \\
& & 0 & 1 & & \\
& & -1& 0 & & \\
& & & & 0 & 1 \\
& & & & -1 & 0
\end{pmatrix}_.
\end{equation}
Each block $M =\begin{pmatrix}
0 & 1 \\
-1 & 0
\end{pmatrix}$ defines a sympletic geometry on a $T^{2}$. It also describes the
intersection matrix of the homology basis. If we denote by $a$ and $b$ the two elements
of the basis of $T^{2}$, $M_{ab}$ is defined by the following intersection numbers: $a.b=-b.a=1$
and $a.a=b.b=0$. For simplicity on our analysis we will take $n=1$, the general case only involve some technical additional details.
\\

The above topological condition leads to a $D=11$ supermembrane with non-trivial central charges
generated by its wrapping on the compact part of the target space. Since the topological
constraint commutes with the rest of the constraints, it represents a sector of the full theory
characterized by an integer $n=det \omega^{rs}$, see~\cite{gmr} for a more general discussion.
Under such correspondence there exists a minimal holomorphic immersion from the base to the target manifold.
The image of $\Sigma$ under that map is a calibrated submanifold of $T^{6}$. The spectrum of the theory changes
dramatically since it has a pure discrete spectrum at the classical and the quantum level~\cite{gmr,bgmmr,bgmr,bgmr2,br};
see also~\cite{joselen,bellorin} \footnote{The geometrical interpretation of this condition has been discussed
in previous work~\cite{torrealba},\cite{mor}}.
\\

The model that we study here involves additional symmetries beyond the original ones~\cite{bgmr2}
which will be crucial in our coming discussion. In the following the minimally immersed M2 associated
to this sector of the theory will be denoted by MIM2 to distinguish it from the usual one.
\\

We notice that the condition in (\ref{e8}) only restricts the values of $S_{s}^{r}$. From equation (\ref{e5})
we can see that these values should be integral numbers. The condition in (\ref{e8}) can be solved by
\begin{equation}
dX^{r}=M_{s}^{r}d\widehat{X}^{s}+dA^{r}
\end{equation}
where we have decomposed the closed one-forms $dX^{r}$ into their harmonic plus exact parts.
 Note that $d\widehat{X}^{s}, s=1,\dots,2g$ is a basis for harmonic one-forms over $\Sigma$.
 They may be normalized with respect to the associated canonical basis of homology,
\begin{equation}
\oint_{c_{s}}d\widehat{X}^{r}=\delta_{s}^{r}.
\end{equation}
We have now considered a Riemann surface with a class of an equivalent canonical basis.
The condition in (\ref{e5}) leads to
\begin{equation}
M_{s}^{r}=2\pi R^{r}S_{s}^{r}.
\end{equation}
Imposing the condition in (\ref{e8}), we get
\begin{equation}
S_{t}^{r}\omega^{tu}S_{u}^{s}=\omega^{rs},
\end{equation}
which says that $S\in Sp(2g,Z)$. This is the most general map satisfying (\ref{e8}).
\\

A sufficient condition in order to have a consistent global construction of the
theory, subject to the topological constraint, is to have a surface $\Sigma$ of genus $g$
such that the space of holomorphic one-forms is of the same complex dimension as the flat
torus in the target space. This condition ensures the existence of a holomorphic immersion,
and so minimal, from $\Sigma$ to $T^{2g}$~\cite{bellorin}. In~\cite{joselen} we analyzed the
theory for genus $3$ and the breaking of the SUSY by the ground state (the holomorphic immersion)
for genus $1,2,3$. It was also emphasized there that in order to consider the MIM2 from $\Sigma$
to a given target space one should consider all possible immersions, in particular all holomorphic
immersions. This consideration will become important in the following sections when we analyse a $\frac{T^{7}}{Z_{2}^{3}}$ target space.
\\

The theory is invariant not only under the diffeomorphisms generated by $\phi_{1}$ and
$\phi_{2}$ but also under the diffeomorphisms, which are biholomorphic maps, changing
the canonical basis of homology by a modular transformation.
\\

We may always consider a canonical basis such that
\begin{equation}\label{e11}
dX^{r}=2\pi R^{r}d\widehat{X^{r}}+dA^{r}.
\end{equation}
In this manner, the corresponding degrees of freedom are described exactly
 by the single-valued fields $A^{r}$. By using the condition in (\ref{e6}),
 we perform a similar decomposition with the remaining 1-form associated to the compactification on $S^{1}$
\begin{equation}\label{e10}
dX^{7}=2\pi R L_{s}d\widehat{X}^{s}+d\widehat{\phi}
\end{equation}
where $d\widehat{\phi}$ is a new exact 1-form and $d\widehat{X}^{s}$
are the basis of harmonic forms as before. The final expression of
the hamiltonian of the MIM2 wrapped in an irreducible way on
$T^{6}\times S^{1}$~\cite{joselen} is
\begin{equation}\label{e}
\begin{aligned}
H=&\int_{\Sigma} \sqrt{w}d\sigma^{1}\wedge
d\sigma^{2}[\frac{1}{2}(\frac{P_{m}}{\sqrt{W}})^{2}
+\frac{1}{2}(\frac{\Pi^{r}}{\sqrt{W}})^{2}+\frac{1}{4}\{X^{m},X^{n}\}^{2}+\frac{1}{2}(\mathcal{D}_{r}X^{m})^{2}\\
&
+\frac{1}{4}(\mathcal{F}_{rs})^{2}+\frac{1}{2}(F_{ab}\frac{\epsilon^{ab}}{\sqrt{W}})^{2}
+\frac{1}{8}(\frac{\Pi^{c}}{\sqrt{W}}\partial_{c}X^{m})^{2}+\frac{1}{8}[\Pi^{c}\partial_{c}(\widehat{X}_{r}+A_{r})]^{2}]\\
&  +
\Lambda(\{\frac{P_{m}}{\sqrt{W}},X^{m}\}-\mathcal{D}_{r}(\frac{\Pi^{r}}{\sqrt{W}})
-\frac{\Pi^{c}}{2\sqrt{W}}\partial_{c}(F_{ab}\frac{\epsilon^{ab}}{\sqrt{W}}))+\lambda\partial_{c}\Pi^{c}]\\
&+\int_{\Sigma}
\sqrt{W}[-\overline{\Psi}\Gamma_{-}\Gamma_{r}\mathcal{D}_{r}\Psi +
\overline
\Gamma_{-}\Gamma_{m}\{X^{m},\Psi\}+1/2\overline{\Psi}\Gamma_{7}\Pi^{b}\partial_{b}\Psi]+\Lambda
\{\overline{\Psi}\Gamma_{-}, \Psi\}
\end{aligned}
\end{equation}
\newline
where $\mathcal{D}_r X^{m}=D_{r}X^{m} +\{A_{r},X^{m}\}$,
$\mathcal{F}_{rs}=D_{r}A_s-D_{s }A_r+\{A_r,A_s\}$ and \\ $D_{r}=2\pi
R^{r}\frac{\epsilon^{ab}}{\sqrt{W}}\partial_{a}\widehat{X}^{r}\partial_{b}$.
$P_{m}$ and $\Pi_{r}$ are the conjugate momenta to $X^{m}$ and
$A_{r}$ respectively. $\mathcal{D}_{r}$ and $\mathcal{F}_{rs}$ are
the covariant derivative and curvature of a symplectic
noncommutative theory~\cite{mor,bgmmr}, constructed from the
symplectic structure $\frac{\epsilon^{ab}}{\sqrt{W}}$ introduced by
the central charge. The physical degrees of the theory are then
described by $X^{m}, A_{r}$, and the corresponding spinorial ones
$\Psi_{\alpha}$. They are single valued fields on $\Sigma$.
\\

At this level, one might naturally ask the following question.
Does there exist a MIM2 compactified on a seven dimensional manifold with G2
holonomy group? In what follows we address this question using a recent result
from algebraic geometry of toroidal compactification in the presence of discrete symmetries.
\section{G2 compactification in M-theory}
As we mentioned in the introduction, a possible way to get four dimensional models
with four supercharges is to consider the compactification of M-theory on seven dimensional
manifolds with G2 holonomy group\footnote{G2 is a group of dimension 14 and rank 2.}~\cite{ABLY,BG,RRW,santillan}.
We will refer to them as G2 manifolds. In this manner, different $N = 1$ models in four dimensions depend
on the geometric realization of the G2 manifold. As for the Calabi-Yau case, there are many geometric realizations.
In what follows we quote some of them~\cite{J}.
\subsection{G2 manifolds}
Let us consider $R^7$ parametrized by $(x_1, x_2, ...., x_7)$. On this space, one can define the metric
as $g = dx^2_1 + .... +dx^2_7$. Reducing the group $SO(7)$ to G2, there is a special real three-form
\begin{equation}\label{3form}
\Psi=dx_{127}+dx_{135}-dx_{146}-dx_{236}-dx_{245}+dx_{347}+dx_{567}
\end{equation}
where $dx_{ijk}$ denotes the exterior form $dx_i\wedge dx_j\wedge dx_k$.
This expression for $\Psi$ arises from the fact that G2 is the group of automorphisms for the octonionic algebra structure given by
\begin{equation}
t_it_j =- \delta_{ij} + f_{kij}t_k
\end{equation}
which yields the correspondence
\begin{equation}
f_{k ij} \to dx_{kij}.
\end{equation}
In general if a seven Riemanian metric admits a covariant constant spinor the holonomy group is G2 and there is exactly one.
In such manifolds there exists an orthogonal frame, $\widehat{e}^{i}$, in which the octonionic
three form $\phi=f_{ijk}\widehat{e}^{i}\wedge\widehat{e}^{j}\wedge \widehat{e}^{k}$ and its dual are
closed. $\phi$ is G2 invariant. It turns out that the simplest example of G2 manifolds, which we are
interested in here, is the orbifold realization. Let us consider a 7-tori $T^7=R^7/Z^7$, where now $x$ parameterizes
$R/Z$. A G2 manifold can be constructed from an orbifold action $T^7/\Gamma$ where $\Gamma$ a discrete subgroup of G2,
and hence leaving the above three-form $\Psi$ invariant. A possible choice is given by
\begin{equation}
\Gamma=Z_2\times Z_2\times Z_2
\end{equation}
to be defined in the next section.
\subsection{ $Z_{2}\times Z_{2}\times Z_{2}$ symmetries of the G2 structure}\label{Z2xZ2xZ2}
The $Z_{2}$ symmetries leaving invariant the 3-form ~(\ref{3form}), which we will consider,
change signs on certain elements of the basis for the octonions. A change of sign for one element
of the basis condemns the same for other elements. These combinations are given by the multiplication
table. For convenience in further identifications we have chosen the multiplication table represented in
Figure~1, where the $e_i$ are the elements from the basis of the octonions. The result for the multiplication
of two elements in the basis is the only other element that shares the line passing through the first two, and
the sign is given by the arrows. For example, $e_6e_7=e_5$ while $e_5e_2=-e_4$.
\\
\begin{figure}[h]
\begin{center}
\includegraphics[height=5cm]{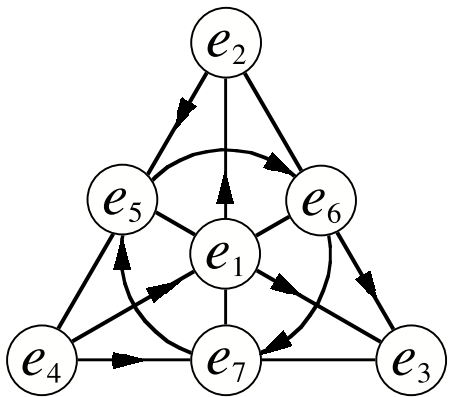}
\vspace*{-0.4cm}
\caption{{\protect\small Fano plane representing the multiplication table for the octonions used throughout this paper.}}
\end{center}
\end{figure}
\\[14.5pt]
A very quick way to determine such subsets of elements is by considering the canonical quaternionic subspaces of the octonions.
\\[14.5pt]
\begin{figure}[h]
\begin{center}
\includegraphics[height=3cm]{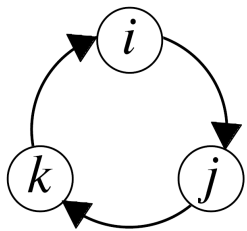}
\vspace*{-0.4cm}
\caption{{\protect\small quaternionic diagram.}}
\end{center}
\end{figure}
\\
Changing signs for the elements in these subsets or their complements each preserve the octonionic structure.
The former maps $\Psi\to-\Psi$ while the latter leaves $\Psi$ completely unchanged. According to the multiplication
table we have chosen, the indexes of the elements from the basis corresponding to these sets are given as follows:
\begin{table}[h]
\begin{center}
\begin{tabular}{|l|l|l|l|l|l|l|l|}
\hline
$\Psi \to -\Psi$ & 1,2,7 & 1,3,5 & 1,4,6 & 2,3,6 & 2,4,5 & 3,4,7 & 5,6,7 \\
\hline
$\Psi \to \Psi$ & 3,4,5,6 & 2,4,6,7 & 2,3,5,7 & 1,4,5,7 & 1,3,6,7 & 1,2,5,6 & 1,2,3,4\\
\hline
\end{tabular}
\caption{Transformations that preserve the octonionic structure.}
\end{center}
\end{table}
These seven transformations obtained by changing signs for the elements on the second file,
together with the identity, form a commutative group with eight elements of order two. This group
is ${Z}_2\times{Z}_2\times{Z}_2 \cong {Z}_{2}^{3}$. There is a nice geometric interpretation for the
operation in this group. Given two transformations, they correspond to two quaternionic subspaces of
the multiplication table for the octonions and share only one element -see the first row in the previous table.
The composition of these transformations is the one related to the only other quaternionic subspace that shares
this element in common. Using the same labeling for the multiplication table of the octonions as the one that
determines~(\ref{3form}) we can list all the ${Z}_2$ symmetries that leave invariant the 3-form $\Psi$ as follows,
\begin{table}[h]
\begin{center}
\begin{tabular}{|ccc|}
\hline
elements that change sign&&the element in ${Z}_2^3$\\
\hline
$x_3,x_4,x_5,x_6$&$\longleftrightarrow$&(0,1,1)\\
$x_2,x_4,x_6,x_7$&$\longleftrightarrow$&(1,1,1)\\
$x_2,x_3,x_5,x_7$&$\longleftrightarrow$&(1,0,0)\\
$x_1,x_4,x_5,x_7$&$\longleftrightarrow$&(0,1,0)\\
$x_1,x_3,x_6,x_7$&$\longleftrightarrow$&(0,0,1)\\
$x_1,x_2,x_5,x_6$&$\longleftrightarrow$&(1,0,1)\\
$x_1,x_2,x_3,x_4$&$\longleftrightarrow$&(1,1,0)\\
\hline
\end{tabular}
\\
\caption{$Z_2^3$ transformations preserving G2 structure.}
\end{center}
\end{table}
and, naturally, the identity transformation is in correspondence with (0,0,0).
\\
Aiming towards a $T^2\times T^2\times T^2\times S^1$ compact space,
we shall identify the coordinates $(x_1,x_2,x_3,x_4,x_5,x_6,x_7)$
with $(z_1,z_2,z_3,x_7)\in{\mathbb C}\times{\mathbb C}\times{\mathbb
C}\times{\mathbb R}$ writing $z_k=x_{2k-1}+ix_{2k}$ for $k=1,2,3$.
The transformations given in the previous table are then expressed
as in Table 3.
\begin{table}[h]
\begin{center}
\begin{tabular}{|lc|cc|}
\hline
\hspace*{20pt}symmetry transformation&\hspace*{0pt}&\hspace*{0pt}&the element in ${Z}_2^3$\\
\hline
$(z_1,z_2,z_3,x_7)\to(z_1,-z_2,-z_3,x_7)$&&&(0,1,1)\\
$(z_1,z_2,z_3,x_7)\to(\bar{z_1},\bar{z_2},\bar{z_3},-x_7)$&&&(1,1,1)\\
$(z_1,z_2,z_3,x_7)\to(\bar{z_1},-\bar{z_2},-\bar{z_3},-x_7)$&&&(1,0,0)\\
$(z_1,z_2,z_3,x_7)\to(-\bar{z_1},\bar{z_2},-\bar{z_3},-x_7)$&&&(0,1,0)\\
$(z_1,z_2,z_3,x_7)\to(-\bar{z_1},-\bar{z_2},\bar{z_3},-x_7)$&&&(0,0,1)\\
$(z_1,z_2,z_3,x_7)\to(-z_1,z_2,-z_3,x_7)$&&&(1,0,1)\\
$(z_1,z_2,z_3,x_7)\to(-z_1,-z_2,z_3,x_7)$&&&(1,1,0)\\
\hline
\end{tabular}
\caption{}
\end{center}
\end{table}
All these symmetries can be obtained as composition of the three
canonical generators, $(1,0,0)$, $(0,1,0)$, and $(0,0,1)$, for
${Z}_2^3$. Nevertheless, there are 28 different subsets of
generators for ${Z}_2^3$ but all geometrically equivalent.
\section{ MIM2 on a G2 manifold}
In this section we will consider the construction of a MIM2 on a $G_{2}$ manifold.
 We start from the MIM2 on a seven torus $T^{7}$ and we will perform
 the identification of the $Z_{2}\times Z_{2}\times Z_{2}$ group, described
 in Section~3, on the target space.
\\

The MIM2 theory on $T^{7}$ is invariant under the area preserving
diffeomorphisms. The ones homotopic to the identity are generated by
the area preserving constraints (\ref{e2}) and (\ref{e3}). The
theory is also invariant under large area preserving
diffeomorphisms, non-homotopic to the identity, associated to
$Sp(6,Z)$ acting on a Teichm\"uler space of the moduli space of
$g=3$ Riemann surfaces as explained in Section~2. We will now show
that the $Z_{2}\times Z_{2}\times Z_{2}$ automorphisms of
quaternionic subspaces of the octonionic algebra described in
Section~3.2 are also symmetries of the hamiltonian of the MIM2 on
$T^{7}$. Moreover, those are the maximal identifications we can
perform on the target space preserving $N=1$ susy. We will see that
the remaining symmetries of the fiber become spureous whenever the
orbifold action on the states is considered.
\subsection{Minimal immersions on the target space}
The maps (\ref{e11},\ref{e10}) from the base $\Sigma$ (g=3) to the
compact sector of the target space $T^{7}$ decompose into a harmonic
part plus an exact one. The harmonic part is a minimal immersion
from $\Sigma$ to the compact sector $T^7$ of the target space. The
requirement introduced in~\cite{joselen} was to consider all
possible immersions from the base manifold to the target space. It
has a natural interpretation in terms of the existence of fluxes on
the
 compact sector of the target space. In fact, the existence of fluxes is equivalent to the existence of a
 bundle gerbe or higher order bundle on the target space\cite{giraud}, \cite{brylinski}\cite{caicedo},
 and \cite{baez1},\cite{baez2}, \cite{isidro},\cite{mackaay}. Given a closed p-form $W_{p}$ satisfying the
 quantization condition
\begin{equation}\label{ew}
\int_{\Sigma_{p}}W_{p}=2\pi n
\end{equation}
for any $\Sigma_{p}$ submanifold there always exists a bundle gerbe
or higher order bundle with its corresponding transition functions
on $p-1,\dots, 1$ forms such that $W_{p}$ is the field strength of a
generalized connection. The consistency condition on the transition
functions is now satisfied on the overlapping of $p+1$ open sets of
an atlas. For the case $p=2$, it is a $U(1)$ principle bundle and
the quantization condition ensures the existence of a connection on
it such that $W_{2}$ is its curvature. The condition (\ref{ew}) must
be satisfied for all $\Sigma_{p}$ submanifolds, the integer $n$ may
change with $\Sigma_p$. If we interpret the central charge condition
as a flux condition on the target, we must then impose it for all
admissible minimal immersion form $\Sigma$ to $T^7$.  In the case of
the MIM2 on a $T^{7}$ target we should then consider all possible
immersions and impose for each of them the topological or central
charge condition. This is a geometrical argument emphasizing that we
should consider the summation of all possible immersions from
$\Sigma$ to the target, see also \cite{zehra}.
\\

We now proceed to consider all possible immersions from
$\Sigma$, a genus 3 Riemann surface to $T^{7}=S^{1}\times\dots\times S^{1}$.
 The reason to consider a genus 3 surface was explained in Section~2, they are
  the relevant ones when considering the wrapping of a supermembrane on a $T^{6}$ target.
  We consider all decompositions of $T^{7}$ into $T^{6}\times S^{1}$, by changing
  the $S^{1}$ we obtain the complete set of seven sectors. The supermembrane
  wraps in an irreducible way onto the $T^{6}$, we ensure it by imposing the topological
  condition on all configurations of the supermembrane on that sector. We distinguish each
  sector by an integer $i=1,\dots,7$ and denote the corresponding maps to the $T^{6}$ by $X_{i}^{r}$, $r=1,\dots,6$
\begin{equation}\label{eA}
dX^{r}=2\pi R S_{is}^{r}d\widehat{X}^{s}+ dA^{r}
\end{equation}
while the remaining one to the $S^{1}$ by $X$
\begin{equation}\label{eB}
dX=2\pi R m_{ir}d\widehat{X}^{r}+dA
\end{equation}
where $S_{is}^{r}\in Sp(6,Z)$ for each i=1,\dots,7, $dA^{r}$ and
$dA$ are exact one-forms. These ones are completely general without
restrictions  as well as the spinor fields on the target which are
also scalars on the worldvolume. They carry the local degrees of
freedom of the supermembrane. For each $T^{6}$ we provide a
symplectic structure in order to define the topological condition on
Section~2, they are given in Table 4.

We will denote by $\Gamma\equiv Z_{2}^{3}$ the discrete group whose
elements change the sign of the maps from $\Sigma$ to $T^{7}$
according to the second row in Table~1. We will denote by
$\Lambda\equiv Z_{2}^{4}$ the discrete group whose elements change
the sign of the maps from $\Sigma$ to $T^{7}$ according to the
complete Table~1. $\Gamma$ is a discrete subgroup of G2 and
$\Lambda$. For each sector $i$ we may associate a subgroup
$Z_2\times Z_2$ of $\Gamma$ in the following way. Take on the first
row of Table 1, the triplets containing an integer $i$, for example
if $i=7$: $(1,2,7), (4,3,7), (5,6,7)$. The corresponding elements on
the second row of Table 1:  $(3,4,5,6),(1,2,5,6),(1,2,3,4)$
respectively determine a subgroup $Z_2\times Z_2$ of $\Gamma$. These
transformations map the sector $i$ into itself. They belong  to the
$Sp(6,Z)$ associated to the sector. We will now show that the other
elements of $\Gamma$ transform admissible maps (the ones satisfying
the topological constraint) of one sector into admissible maps of
another one. The integers $m_{ir}$ become determined in the
procedure.

\paragraph{Computation of $m_{ir}$}
We start with the most general expression (\ref{eA}), (\ref{eB}) in
sector $i=7$, by performing a change on the homology basis and in
the corresponding normalized basis of one-forms, it can always be
reduced to,
\begin{equation}
d\widehat{X}^{1},d\widehat{X}^{2},d\widehat{X}^{3},d\widehat{X}^{4},d\widehat{X}^{5},d\widehat{X}^{6},m_{7r}d\widehat{X}^{7}
\end{equation}
where from now on we denote in a file the harmonic part of $dX^{i}$,
for each $i$ ordered from 1 to 7. To simplify the notation we do not
write explicitly the $2\pi R$ factors. The exact part is not
relevant in the determination of the admissibility of a map, and may
be added at any stage of the argument. If we now apply
transformation $(2,4,6,7)$ the new map
\begin{equation}\label{eC}
d\widehat{X}^{1},-d\widehat{X}^{2},d\widehat{X}^{3},-d\widehat{X}^{4},d\widehat{X}^{5},-d\widehat{X}^{6},-m_{7r}d\widehat{X}^{7}
\end{equation}
is not admissible in sector 7 but it is in the other sectors. For example if we take the sector 1,
with symplectic structure given in Table~4 below, it is admissible if
\begin{equation}
m_{7r} d\widehat{X}^{r}=d\widehat{X}^{1}+m_{72}d\widehat{X}^{2}
\end{equation}
for any integer $m_{72}$.
\\

If we now consider the transformation $(2,3,5,7)$ of $\Gamma$, (\ref{eC}) transforms into
\begin{equation}
d\widehat{X}^{1},-d\widehat{X}^{2},-d\widehat{X}^{3},d\widehat{X}^{4},-d\widehat{X}^{5},d\widehat{X}^{6},-m_{7r}d\widehat{X}^{7}
\end{equation}
which is admissible in sector 1 for any $m_{72}$. Under $(1,4,5,7)$ (\ref{eC}) transforms into
\begin{equation}
-d\widehat{X}^{1},d\widehat{X}^{2},d\widehat{X}^{3},-d\widehat{X}^{4},-d\widehat{X}^{5},d\widehat{X}^{6},-d\widehat{x}^{1}-m_{7r}d\widehat{X}^{7}
\end{equation}
it is admissible only in sector 2 with $m_{72}=1$. Finally under $(1,3,6,7)$ (\ref{eC}) transforms into
\begin{equation}
-d\widehat{X}^{1},d\widehat{X}^{2},-d\widehat{X}^{3},d\widehat{X}^{4},d\widehat{X}^{5},-d\widehat{X}^{6},-d\widehat{x}^{1}-d\widehat{X}^{2}m_{7r}
\end{equation}
which is also admissible in sector 2. The general values of $m_{7r}$ in order to have the full $\Gamma$ as a symmetry on the admissible set of maps is
\begin{equation}\label{eE}
%\begin{eqnarray*}
m_{7r}d\widehat{X}^{r}=
\begin{cases}
\pm (d\widehat{X}^{1}+d\widehat{X}^{2})\\
\pm (d\widehat{X}^{3}+d\widehat{X}^{4})\\
\pm (d\widehat{X}^{5}+d\widehat{X}^{6})
\end{cases}
%\end{eqnarray*}
\end{equation}
The general expression for $m_{ir}$ is obtained from $m_{7r}$ by applying the elements of $\Gamma$.
\\

We conclude then that given a general admissible map on any sector
there always exists another admissible map which is the transformed
under $\Gamma$ of the original one. The integers $m_{r}$ take some
particular value in the procedure. In other words, for that
particular values of $m_{r}$, the set of admissible maps is
preserved under the action of $\Gamma$. Moreover, the bosonic
hamiltonian as a map from the space of configurations to the reals
is invariant under $\Gamma$. The same properties are valid for the
discrete group $\Lambda=Z_{2}^{4}$, with the same values of $m_{r}$.
However, we are considering the wrapping of the MIM2 on an oriented
$T^7$ and the transformations on the first row of Table 1. do not
preserve the orientation of $T^7$. We are then left only with the
group of discrete symmetries $\Gamma=(Z_2^3)$. The supersymmetric
hamiltonian is invariant under these symmetries. All other discrete
symmetries of the hamiltonian whose bosonic part is quartic but
quadratic on each map, are not symmetries of the admissible set.
\\

The symplectic structure on each sector we have used is given in Table~4. We notice that there is no
loss of generality by using it, since on any other election of the symplectic matrices
the above properties of the admissible set are also valid. The only change is on
the explicit realization of the maps,
\begin{table}[h]
\begin{center}
\begin{tabular}{|rcl|}
\hline
$\omega_{7}$&=&$dX^{1}\wedge dX^{2}+dX^{3}\wedge dX^{4}+dX^{5}\wedge dX^{6}$\\
$\omega_{6}$&=&$dX^{2}\wedge dX^{1}+dX^{4}\wedge dX^{3}+dX^{5}\wedge dX^{7}$\\
$\omega_{5}$&=&$dX^{2}\wedge dX^{1}+dX^{4}\wedge dX^{3}+dX^{7}\wedge dX^{6}$\\
$\omega_{4}$&=&$dX^{2}\wedge dX^{1}+dX^{3}\wedge dX^{7}+dX^{6}\wedge dX^{5}$\\
$\omega_{3}$&=&$dX^{2}\wedge dX^{1}+dX^{7}\wedge dX^{4}+dX^{6}\wedge dX^{5}$\\
$\omega_{2}$&=&$dX^{1}\wedge dX^{7}+dX^{4}\wedge dX^{3}+dX^{6}\wedge dX^{5}$\\
$\omega_{1}$&=&$dX^{7}\wedge dX^{2}+dX^{4}\wedge dX^{3}+dX^{6}\wedge dX^{5}$\\
\hline
\end{tabular}
\\
\caption{}
\end{center}
\end{table}

\paragraph{Remark}
It is important to emphasize the relation between $\Gamma$ and the $Sp(6,Z)$ group of large area
 preserving diffeomorphisms. The space of admissible maps is invariant under the full group $Sp(6,Z)$.
It transforms admissible maps of one sector into admissible maps of the same sector. Its action on the harmonic
 sector of the maps shares in common with $\Gamma$ a subgroup $Z_{2}\times Z_{2}$ provided
 the maps are restricted to that sector. When the $Z_{2}\times Z_{2}$ is lifted to the whole admissible set it is not anymore a
transformation generated by an area preserving diffeomorphism. However it is a symmetry, as
we have shown, of the admissible set.
$\Gamma$ is then the unique discrete symmetry relating the different sectors of the admissible set;
the $Sp(6,Z)$ acts only on each sector. We notice that these sectors arise from the different possible
wrappings of the MIM2 on $T^{7}$ and their origin is not related to the twisted or untwisted
sectors of the MIM2 when the identification on an orbifold is performed.

\subsection{Configuration space}
We will now define a MIM2 on the $G2$ orbifold $T^{7}/ \Gamma$
constructed by Joyce~\cite{J}. The group of transformations
$\Gamma=Z_{2}^{3}$ introduced by Joyce has additional shifts with
respect to the transformations in Section~3.2. Those shifts are
irrelevant concerning the action of the group in the MIM2 theory
since the maps only enter in terms of one-forms and hence the shifts
disappear. However they are important in the construction of the
orbifold. These shifts can be generated in the MIM2 theory by the
constraint (\ref{e3}). It generates area preserving diffeomorphisms
homotopic to the identity with infinitesimal parameters which are
harmonic one-forms on $\Sigma$. The transformation for the maps is
\begin{equation}\label{ex} \delta X^{s}=\{X^{s},\xi\} \end{equation} with
$\xi=\xi_{r}\widehat{X}^{r}$, $\quad r=1,\dots,6$. The harmonic part
is then shifted by (\ref{ex})
\begin{equation} \frac{1}{Area_{\Sigma}}\int_{\Sigma}\delta X^{r}\sqrt{W}d\sigma^{1}\wedge
d\sigma^{2}=\omega^{rs}\xi_{s}. \end{equation} We may then fix six
shifts, corresponding to the mean value of the map over $\Sigma$. In
the notation of Section~3.2 the generators of the Joyce $Z_{2}^{3}$
are: $\alpha=(3,4,5,6), \beta= (1,2,5,6), \gamma=(2,4,6,7)$ with the
same shifts of value $1/2$. That is,
\begin{equation}\begin{aligned} \alpha: (x^{1},\dots,x^{7})&\to &
(x^{1},x^{2},-x^{3},-x^{4},-x^{5},-x^{6},x^{7})\\ \nonumber \beta:
(x^{1},\dots,x^{7})&\to &
(-x^{1},-x^{2},x^{3},x^{4},1/2-x^{5},-x^{6},x^{7})\\ \nonumber
\gamma: (x^{1},\dots,x^{7})&\to &
(x^{1},x^{2},-x^{3},1/2-x^{4},-x^{5},1/2-x^{6},x^{7}).\\ \nonumber
\end{aligned}\end{equation} The elements of the group $\Gamma$ are isometries of $T^{7}$,
preserving its flat $G_{2}$-structure. The fixed points of
$\alpha,\beta,\gamma$ are each 16 copies of $T^{3}$. The singular
set $S$ of $\frac{T^{7}}{\Gamma}$ is a disjoint union of 12 copies
of $T^{3}$. The singularity on each component of $S$ is of the form
$T^{3}\times \frac{\mathbb{C}}{\pm 1}$. The singularities of
$T^{7}/\Gamma$ can be resolved and a metric with holonomy G2 on a
compact 7 manifold may be obtained~\cite{J}.
\\
\paragraph{Untwisted Sector}

We may now consider the construction of the untwisted sector of the
MIM2 on the G2 orbifold $T^{7}/\Gamma$. We start from the general
space of configurations satisfying the topological constraint
ensuring the irreducible wrapping of all configurations of the
supermembrane. We then consider the subspace of configurations
invariant under $\Gamma$. This was constructed in section 4.1. The
maps are of the form (\ref{eA}) (\ref{eB}) with the restrictions on
the values of $m_{ir}$ (\ref{eE}) (on that particular basis of
harmonic one-forms). On the space of configurations we construct
classes, two elements of a class are related by a transformation of
$\Gamma$. The hamiltonian, as mentioned before has the same value on
each element of the class. The untwisted sector of the theory is now
defined on the space of classes. Each class represents now the map
from $\Sigma$ to the orbifold. This construction may be implemented
directly in the functional integral of the supermembrane, which in
the case of the MIM2 has a well defined gaussian measure. The
untwisted sector we have constructed breaks SUSY to $N=1$ and it is
directly related to the analysis in \cite{joselen}.
\paragraph{Twisted Sector}
We will denote by $Z_2$ spin structure on a n-dimensional vector
bundle $E$ a principle spin-bundle $P_{Spin}(E)$ together with a
two-sheeted covering: \begin{equation} \xi: P_{Spin}(E)\to P_{SO}(E)
\end{equation} such that $\xi(pg)=\xi(p)\xi_{0}(g)$ for all $p\in P_{Spin}(E)$
and all $g\in Spin$. $\xi_{0}$ is the universal covering
homomorphism $\xi_{0}: Spin (n)\to SO(n)$ with kernel $\{1,-1\}\sim
Z_{2}$. An element of $P_{SO}(E)$ can be lifted to $P_{Spin}(E)$ if
and only if $W_{2}(p)=0$, where $W_2$ is the second Stiefel-Whitney
class. When $n=1$, $P_{SO}(E)=X$ the base manifold and a spin
structure is defined to be a 2-fold covering of $X$. The $Z_2$ spin
structures when they exist are are in one to one correspondence to a
$+$ or a $-$ assign to the elements of a basis of homology on $X$.

We consider now the construction of the twisted sector of the
M2\footnote{A former study of the twisted states of an extended
membrane in the case of M theory on an orbifold $\frac{S^1}{Z_{2}}$
was considered in \cite{hw}.} with central charges on a G2 manifold.
The group of identifications on the target torus is
$\Gamma=Z_{2}^3$. The twisted sectors correspond to maps which
change sign when going around a cycle on $\Sigma$ according to some
element of $\Gamma$. To construct all the global objects satisfying
such conditions, we proceed as follows. We assign to each element of
the basis of homology $\mathcal{C}_{r}, r=1,\dots, 2g$ an element
$\Gamma_r$ of $Z_2^3$. Each assignment defines a $Z_{2}^3$ spin
structure on the Riemann surface. For such spin structure we
construct the following global object. The map $X^{i}, i=1,\dots, 7$
is a section of $P_{Spin}(X)$, which is a 2-fold covering of $X$,
with a $Z_2$ Spin structure determined by the $+$ or $-$ sign
assigned to the homology basis according to the $i^{th}$ sign $\pm$
associated to the maps $\Gamma_r$. For example let us consider the
$Z_2^3$ Spin-structure obtained by assigning $\Gamma_1=(2,3,5,7)$ to
$\mathcal{C}_1$, $\Gamma_2=(1,4,5,7)$ to $\mathcal{C}_2$ and the
identity $\mathbb{I}$ to the rest $\mathcal{C}_t, t=3,4,5,6$. We
then have the corresponding transformations
\begin{table}[h]
\begin{center}
\begin{tabular}{|l|l|l|l|l|l|l|l|}
\hline
  & $X^{1}$ & $X^{2}$ & $X^{3}$ & $X^{4}$ & $X^{5}$ & $X^{6}$ & $X^{7}$ \\
\hline
$\mathcal{C}_1 \to \Gamma_1$ & +& -& -& +& -& +& -\\
\hline
$\mathcal{C}_2 \to \Gamma_2$ & -& +& +& -& -& +& -\\
$\mathcal{C}_3\to\mathbb{I}$ & +& +& +& +& +& +& +\\
\vdots &\vdots&\vdots&\vdots&\vdots&\vdots&\vdots&\vdots\\
$\mathcal{C}_6\to\mathbb{I}$ & +& +& +& +& +& +& +\\
\hline
\end{tabular}
\caption{The columns of the table define the sections with which the maps of the twisted sector are constructed.}
%{\protect\small Table 5.}
\end{center}
\end{table}

We now construct the global object $X^i,i=1,\dots,7$ by considering
a section of $P_{Spin}$, a 2-covering of the Riemann surface, with
the $Z_2$ spin structure obtained by the columns in the above
diagramme. The corresponding sections may be explicitly constructed
in terms of the harmonic one-forms
$d\widehat{X}^1,\dots,d\widehat{X}^6$ of the $g=3$ Riemann surface:
\begin{equation}\begin{aligned} &X^1=e^{\frac{i}{2}\widehat{X}^2}\varphi^1,\quad
X^2=e^{\frac{i}{2}\widehat{X}^1}\varphi^2,\quad
X^3=e^{\frac{i}{2}\widehat{X}^1}\varphi^1,\quad
X^4=e^{\frac{i}{2}\widehat{X}^2}\varphi^2,\\& \nonumber
X^5=e^{\frac{i}{2}\widehat{X}^1}e^{\frac{i}{2}\widehat{X}^2}\varphi^5,\quad
X^6=\varphi^6,\quad
X^7=e^{\frac{i}{2}\widehat{X}^1}e^{\frac{i}{2}\widehat{X}^2}\varphi^7,\end{aligned}
\end{equation} where $\varphi^r$, $r=1,\dots,6$ are scalar fields wrapping $T^7$ as described in section 4.
 These maps are scalar fields on the 2-fold coverings of the
base Riemann surface $\Sigma$. The space of those maps, for all
possible assignment of elements of $Z_2^3$ to the homology basis
define the twisted sector of the MIM2 theory. They remain being
scalar fields, as required by the supermembrane lagrangian, but are
defined on 2-fold coverings of $\Sigma$.
\paragraph{Remark}
The $Sp(6,Z)$ symmetry on the admissible set is broken after
identifying the points on $T^{7}$ by $\Gamma$. On each sector
of the admissible set one is left with a $Z_{2}\times Z_{2}$ symmetry.
\subsection{Connection with Calabi-Yau compactifications}
It is very well known that the G2 manifold can be also built using a partial complex
 structure coordinate~\cite{RRW}. The above 3-form can be re-expressed as
\begin{equation}\label{e1}
\Psi= Re(\Omega)+w\wedge dx_7.
\end{equation}
In this equation $\Omega=dz_1\wedge dz_2\wedge dz_3$ is the complex holomorphic
form of $C^3$ and $w=\frac{i}{2}(dz_1\wedge d{\bar z_1}+dz_2\wedge d{\bar z_2}+dz_3\wedge d{\bar z_3})$
is the Kahler form. Since SU(3) is a subgroup of G2, one can identify the $C^3$ factor with a local
Calabi-Yau threefold (CY3) used in two dimensional $N=2$ sigma model~\cite{Ws,AV,Be}.
In this realization, the above three-form (\ref{e1}) is invariant under the symmetry
\begin{eqnarray}\label{cyt}
 z_{i}\to \overline{z_{i}} \quad x_{7}\to -x_{7}.
\end{eqnarray}
which is needed to ensure $N=1$ in 4D. We will try to show that this
transformation can be related to the above $Z_2 \times Z_2\times
Z_2$ symmetry used in the orbifold construction. This can be done by
imposing certain constraints depending on the precise $Z_{2}$
action. Indeed, the CY3 could be taken as $T^2\times T^2\times T^2$
quotiented by $Z_2 \times Z_2$. Since the CY condition requires the
use of only two $Z_2$'s ( $Z_2^1 \times Z_2^2$), we need to single
out the third $Z_{2}^{3}$ factor. $Z_2^1\times Z_2^2$ acts on the
six-torus structure, producing as a result a CY3, and trivially on
the circle $S^1$. The third $Z_{2}^{3}$ acts on both, the CY3 and
the circle leading to the G2 structure manifold. In this way, one
can identify the last action with the transformation given in
(\ref{cyt}).
\\

The singularities of this orbifold can be identified with its fixed points.
In the three dimensional complex factor, the fixed locus of this G2 manifold is a Lagrangian
submanifold. Its volume form is defined by the real part of $\Psi$. Since the circle has two
fixed points, the total singular geometry then consists of two copies of such a lagrangian submanifold.
The singularities can have an interpretation in the MIM2 picture as critical points. However this does
not mean that there is a degenerate locus of extremal points. On the contrary, the quantum analysis reveals
that there is an absolute minimum for the hamiltonian of the supermembrane. There are no flat directions
in the potential. This fact can be understood from the fact that the dual of the gauge symmetries correspond
 to different backgrounds and not a unique one.
\\

Locally each singular point should be resolved like $R^3 \times X$, where $X$ is an ALE Calabi–Yau 2-fold
asymptotic to $C^2/{Z_2}$, is known as ALE space with $A_1$ singularity. The ALE space with $A_1$ singularity is described by
\begin{equation}
z^2_1 + z^2_ 2 + z^2_3 = 0.
\end{equation}
Using a simple change of variables, this is equivalent to
\begin{equation}
xy = z^2
\end{equation}
where $x$, $y$ and $z$ are complex coordinates. As usual, this singularity can be removed either by deforming
the complex structure or by a blow-up procedure. Geometrically, this corresponds to replacing the singular
point $(x = y = z = 0)$ by a $CP^1 \sim S^2$. As previously explained the (APD) connected with the identity
deform the shape of each $T^2$ and they produce translation on the orbifold side. They serve to blow up the
corresponding orbifold singularities leading to a compactification on a true G2 manifold.
\section{Quantum properties of the supersymmetric theory}
In this section we discuss the quantum consistency of our previous
construction of the supersymmetric action of the supermembrane
(minimally immersed) on a G2 manifold. In particular its spectrum is
discrete. Both results are unique and highly nontrivial from the
supermembrane point of view.\newline

 We denote  the regularized hamiltonian of the supermembrane with
the topological restriction by $H$, its bosonic  part $H_b$ and its
fermionic potential $V_f$, then
                              \begin{equation} H = H_b+V_f\end{equation}.
We can define rigorously the domain of $H_b$ by means of Friederichs
extension techniques. In this domain $H_b$ is self adjoint and it
has a complete set of eigenfunctions  with eigenvalues accumulating
at infinity. The operator multiplication by $V_f$ is relatively
bounded with respect to $H_b$. Consequently  using Kato perturbation
theory  it can be shown that $H$ is self adjoint if we choose
\begin{equation}
Dom{H}=Dom{H_b}\end{equation}.

In \cite{bgmr} it was shown that H possesses a complete set of eigenfunctions and its spectrum is discrete, with finite
multiplicity and with only an accumulation point at infinity. An independent proof was obtained in \cite{br} using the spectral
theorem and theorem 2 of that paper. In section 5 of \cite{br} a rigorous proof of the Feynman formula  for the Hamiltonian of the
 supermembrane was obtained.
In distinction, the hamiltonian of the supermembrane, without the topological restriction, although it is positive, its fermionic
potential is not bounded from below and it is not a relative perturbation of the bosonic hamiltonian. The use of the Lie product
theorem in order to obtain the Feynman path integral is then not justified. It is not known and completely unclear whether a Feynman
path integral formula exists for this case.

In the previous sections we have provided a construction of the supermembrane with the topological restriction on an orbifold with $G_2$
structure that can be ultimately deformed to lead to a true G2 manifold. All the discussion of the symmetries on the Hamiltonian is performed
directly in the Feynman path integral, at the quantum level, and is complete valid by virtue of our previous proofs. All other constructions
in terms of supermembranes  not restricted by  our topological restriction are not justified in any sense and  from a quantum mechanical point
of view probably wrong.

In \cite{bst} action integral of the Supermembrane, the fermionic
fields under the Lorentz transformations on the target space are
scalars under diffeomorphisms on the worldvolume. They are scalars
under area preserving diffeomorphisms, both connected and not
connected to the identity, in the light cone gauge and there is no
harmonic sector related to it. Consequently, it is invariant under
all symmetries introduced  in our construction and the
supersymmetric theory and not only the bosonic part is compactified
on the $G_2$ manifold.
\\
Moreover, in \cite{bgmr2} was proved that the theory of the
supermembrane with central charges, corresponds to a nonperturbative
 quantization of a symplectic Super Yang-Mills in a confined phase and the theory possesses a mass gap.

In distinction with other analysis, the discrete symmetries required
to perform the orbifold with $G_2$ structure identification are
already realized at the level of the hamiltonian leading to a
top-down compactification. This fact restricts the compactification
manifold to a particular one where we can guarantee that all of the
above spectral properties of the supersymmetric hamiltonian
compactified on a torus found before, are preserved on the
compactification
 process on the G2 manifold, for its bosonic and supersymmetric extension,
 which is, a priori, a highly non-trivial fact.
 Indeed, the untwisted sector of the theory is exactly the same
 that the one corresponding to the compactification of the MIM2 on a 7-torus
 with the integers of the minimal immersion in the orbifold case particularized
 to some specific values that do not alter in any sense the spectral properties.
 The twisted sector of the theory only adds a finite number of states compatible with the
 orbifold projection, and it does neither change the spectral properties. On this $G2$ orbifold
 we can guarantee the discreteness of the quantum supersymmetric spectrum of the MIM2.  The deformation
 of the singularities is also due to the invariance of the hamiltonian under the DPA symmetries of the
 former theory, so there is no change in the characterization of the spectrum in the blow-up process. \newline
\section{Phenomenological analysis of the MIM2 on this G2 manifold}
We will show in this section that the model we have exposed above
represent a new kind of models with potential interesting properties
at phenomenological level.

 It has been pointed
out in \cite{AG} the phenomenological interest of G2
compactifications that admit an expression in terms of CY
compactifications since for those manifolds explicit metric can be
obtained , i.e. \cite{jose}, and ALE resolutions of the
singularities may lead to interesting phenomenological properties as
chirality and nonabelian gauge groups, in that sense it is very
appealing to have been able to express our transformations in terms
of that. It has been argued however that orbifold singularities are
not enough to guarantee chirality \cite{acharya-witten}, but it is
needed an isolated conical singularity. Interesting models in which
D6 branes are wrapping Slag cycles of a CY manifold that have an
uplifting in M-theory as  Taub-nut geometry with fractional M2
wrapping collapsed 3-cycles in a G2 compactification can be found in
\cite{edelstein},\cite{mav},\cite{mgu} with interesting
phenomenological properties, also models of M2 on G2
compactifications able to produce non perturbative effects
\cite{moore}. Our approach at first sight could seem to not share
such a nice features, however the study of its phenomenological
properties
 is far to be closed.
We would like to stress that although we have contructed a G2
manifold with orbifold singularities, we have a regular
supermembrane minimally immersed on a G2 and not a fractional one.
As happens in string compactifications there are different ways to
obtain interesting phenomenology: let say Calabi-Yau's
compactifications with Dp branes at the singularities, where the
enhancement of the symmetry is due to the geometry of the
singularity, that has its correspondence with the first type of
models n G2 compactifications \cite{edelstein}-\cite{mgu}. In those
it is fundamental to have a conical singularity on the G2
compactification side. There is a second way to obtain interesting
phenomenology that corresponds to have intersecting Dp branes (IIA)
or magnetized Dp branes (IIB) on, for example, an orientifold
orbifolded action, where the gauge and chiral properties are mainly
due to the particularities of the Dp brane construction. Our
M-theory model  would be
 in correspondence with this second type. Here the chirality properties
and gauge enhancement would be due to the MIM2 worldvolume properties and
not associated to the former orbifolded singularities (that are
smoothed). In that sense it would be
interesting to compute explicitly the corresponding metric and study
its phenomenological properties. Other aspects of interest like
confinement from G2-manifolds \cite{Acharya} ( considered mainly in
G2 manifolds with ALE singularities) emerge naturally in our case
since the spectral properties of the MIM2 have not changed when we
have performed the identification in the target space and the theory shows confinement. In
\cite{bgmr2}, it is argued how the MIM2 theory could reproduce the
strong coupling regime of susy QCD since there are present glueballs
and it possesses a discrete spectrum with a mass gap. Indeed it
corresponds exactly to a symplectic Super Yang-Mills in 4d coupled
to several scalar fields. The proposal is that the confined phase of
the theory corresponds to the MIM2 on a 7-torus and the quark-gluon plasma phase
to the ordinary M2 compactified in a 7-torus. Both phases are
connected through a topological phase transition of quantum origin that breaks the
center of the group. Since the theory of MIM2 on a G2 manifold do
not change its quantum spectral properties, those previous
properties would apply and  it could also described the confined
phase of the theory.
 Regarding moduli stabilization aspects,
assuming the target torus is fixed to be isotropic,
the moduli parametrizing the position of the MIM2 on a
 7-torus as well as the overall moduli parametrizing
the size of the manifold is fixed \cite{joselen}.
When the MIM2 is compactified on the G2 orbifold
the singularities are resolved through
a backreaction effect due to the wrapping, then the moduli
associated to those singularities we argue that they are also fixed.
%\newline
 We have then obtained the 11D supermembrane minimally immersed on a particular G2 manifold.

\section{Discussion and conclusions}
In this paper we have shown, for first time up to our knowledge, a
top-down compactification of the supermembrane on a particular G2
manifold. The 11D supermembrane theory restricted by a topological
condition due to an irreducible wrapping is stable at classical and
quantum level, has discrete spectrum and a mass gap. It can be
compactified on a $\frac{T^{7}}{Z_{2}^{3}}$  orbifold preserving its
quantum stability properties. The resulting theory can be
interpreted as a compactification on a G2 manifold. Indeed, the
symmetries of the theory produce a holonomy bundle that corresponds
exactly to those associated to the Riemanian holonomy of a G2
manifold. By performing the identification on the target space of
the discrete symmetries preserving the topological condition, only
those symmetries associated to the G2 orbifold space are possible,
neither the configuration states nor the minimal immersions are
invariant under the spureous symmetries that would break the
supersymmetry to $N=0$. One can see that the holonomy bundle
associated to the compactification to 5D is related with the Klein
subgroup. When this is further compactified to the remaining $S^{1}$
there exist seven possible immersions of the M2-brane on the target
space of the $T^{7}$ that allow to make exactly the identifications
with the G2 holonomy group. The singularities of this G2 orbifold
may be resolved, as shown by Joyce, leading to a true G2 manifold.
The shifts have their origin in the diffeomorphisms homotopic to the
identity of the MIM2. The untwisted and twisted sector are
completely characterized. Moreover, this result can also be seen in
terms of a $\frac{CY_{3}\times S^{1}}{Z_{2}}$.\newline

We can finally conclude that for first time a consistent
quantization procedure  for the supermembrane on a G2 manifold has
been presented. It is in terms of the supersymmetric action of the
supermembrane, subject to a topological condition -which is
equivalent to have central charges due to an irreducible winding-,
on a particular G2 manifold.\newline

 From a phenomenological point of view, the supermembrane with central charges on the $G_2$ manifold
  represents a new kind of models of compactification in which the supermembrane is minimally immersed a
  long the whole G2 manifold and not just at the singularities. Typically in the literature there has been studied the wrapping
  of M2´s around the singularities of a $G_2$ manifold, in analogy with the constructions of Dp-branes models at singular Calabi-Yau´s
  in string theories. These constructions require particular conditions in order to obtain interesting properties, i.e. chirality is
  associated to the existence of conical singularities on the  $G_2$ manifold and the gauge groups to have orbifold singularities:
  ADE singularities, etc.
In the supermembrane with central charges the gauge field content is already defined on its worldvolume
and is not associated to the singularities of the compactification manifold. Since there is also a flux
condition on the worldvolume, chirality in our model cannot be automatically ruled out -in resemblance with
the magnetized D-brane models on type II constructions- and deserve further study. We  think that the supermembrane
with central charges compactified on this $G_2$ manifold is then an interesting starting point on the construction of a new kind of
models with potentially rich phenomenology.

\section*{Acknowledgements}
We thank L. J. Boya, G. Bonelli, C. Hull, A. Lerda, H. Nicolai, M. Petropoulos,
 N. Prezas, J. Roseel, V. N. Suryanarayana, I. Stavrov,  A. Uranga,
 for helpful discussions. MPGM gives thanks to Departamento de Fisica,
 Universidad de Zaragoza, Spain, for kind invitation and hospitality while
 part of this work was done. AB would like to thank UFR-Lab/ PHE, Rabat for
 hospitality where a part of this work was done, and the project, between
 Universidad de Zaragoza (Spain) and Faculte des Sciences de Rabat (Morocco),
 {\em Fisica de altas energias: Particulas, Cuerdas y Cosmologia} (grant A9335/07).
 MPGM is partially supported by Dipartimento di Fisica di Universita di Torino under
 European Comunity's Human Potential Programme and by the Italian MUR under contracts
 PRIN-2005023102 and PRIN-2005024045. The work of AR is partially supported by a grant
 from MPG, Albert Einstein Institute, Germany and by PROSUL, under contract CNPq 490134/2006-08.
 The work of AB and AS has been supported by CICYT (grant FPA-2006-02315) and DGIID-DGA (grant 2007-E24/2), Spain.

\end{document}